\documentclass[a4paper,12pt]{article}
\usepackage{graphicx}
\usepackage{psfrag}

\usepackage{mathrsfs}
\usepackage{amssymb}
\usepackage{amsfonts}
\usepackage{latexsym}
\usepackage{amsmath}
\usepackage{amsthm}
\usepackage[cp1251]{inputenc}
\usepackage[english]{babel}

\newcommand{\er}[1]{\textrm{(\ref{#1})}}
\def\lb{\label}

\theoremstyle{plain}

\newtheorem{theorem}{\bf Theorem}[section]
\newtheorem{lemma}[theorem]{\bf Lemma}

\theoremstyle{remark}

\newcommand{\g}{\gamma}         \newcommand{\cC}{\mathcal{C}}         
\newcommand{\G}{\Gamma}

\newcommand{\D}{\Delta}                
\newcommand{\ve}{\varepsilon}          
\newcommand{\z}{\zeta}                 
\newcommand{\e}{\eta}                  
\newcommand{\vt}{\vartheta}

\renewcommand{\l}{\lambda}               
               
\newcommand{\m}{\mu}                   
\newcommand{\n}{\nu}                   
\renewcommand{\r}{\rho}                  
\newcommand{\s}{\sigma}           \newcommand{\cR}{\mathcal{R}}

\newcommand{\f}{\phi}                  
                  
\newcommand{\vp}{\varphi}              
                   
\newcommand{\p}{\psi}                   
             \newcommand{\cZ}{\mathcal{Z}}      		
\renewcommand{\o}{\omega}

\newcommand{\x}{\xi}

\newcommand{\vk}{\varkappa}

  \def\mA{{\mathscr A}}
 \def\mB{{\mathscr B}}

  \def\mH{{\mathscr H}}

  \def\mP{{\mathscr P}}

\newcommand{\gD}{\mathfrak{D}}

\newcommand{\gS}{\mathfrak{S}}

\def\J{\mathbb{J}}
\def\Z{\mathbb{Z}}
\def\R{\mathbb{R}}
\def\C{\mathbb{C}}

\def\K{\mathbb{K}}

\def\qqq{\qquad}
\def\qq{\quad}
\let\ge\geqslant
\let\le\leqslant

\newcommand{\ca}{\begin{cases}}
\newcommand{\ac}{\end{cases}}
\newcommand{\ma}{\begin{pmatrix}}
\newcommand{\am}{\end{pmatrix}}

\def\lt{\biggl}
\def\rt{\biggr}

\renewcommand{\[}{\begin{equation}}
\renewcommand{\]}{\end{equation}}
\def\wt{\widetilde}
\def\pa{\partial}
\def\sm{\setminus}
\def\es{\emptyset}
\def\no{\noindent}
\def\ol{\overline}
\def\iy{\infty}
\def\ev{\equiv}
\def\/{\over}
\def\ts{\times}

\def\os{\oplus}
\def\ss{\subset}

\def\Re{\mathop{\rm Re}\nolimits}

\def\Im{\mathop{\rm Im}\nolimits}

\def\BBox{\hspace{1mm}\vrule height6pt width5.5pt depth0pt \hspace{6pt}}

\begin{document}

\title {Effective masses  for zigzag nanotubes
in magnetic fields}

\author{  Evgeny Korotyaev
\begin{footnote}
{Institut f\"ur  Mathematik,  Humboldt Universit\"at zu Berlin,
Rudower Chaussee 25, 12489, Berlin, Germany, e-mail:
evgeny@math.hu-berlin.de }
\end{footnote}
}

\maketitle

\begin{abstract}
We consider the Schr\"odinger operator with a periodic potential on
 quasi-1D models of zigzag single-wall carbon nanotubes in magnetic field. The spectrum of this operator consists of an absolutely continuous part (intervals separated by gaps) plus an infinite number of eigenvalues  with infinite multiplicity.  We obtain
 identities and a priori estimates in terms of effective masses 
 and gap lengths.
\end{abstract}

\vskip 0.25cm
\section {Introduction and main results}
\setcounter{equation}{0}

We consider the Schr\"odinger operator 
$\mH_B =(-i\nabla-\mA)^2+V_q$  with a periodic potential $V_q$ 
on the zigzag nanotube $\G^N\ss\R^3$
(1D models of zigzag single-well carbon nanotubes, see  \cite{Ha}, \cite{SDD}) in a uniform magnetic field $\mB=B(0,0,1)\in \R^3$, $B\in\R$. 
The corresponding vector potential is given by $\mA({\bf x})={1\/2}[\mB,{\bf x}]={B\/2}(-{\bf x_2},{\bf x}_1,0),\  {\bf x}=({\bf x}_1,{\bf x}_2,{\bf x}_3)\in\R^3$. 
Our model nanotube $\G^N$ is a union of edges $\G_\o$ of length 1, i.e.,
$$
\G^N=\cup_{\o\in \cZ} \G_\o,\qq \o=(n,l,j)\in \cZ=\Z\ts \J\ts \Z_N,
\qq \J=\{0,1,2\},\qq \Z_N=\Z/(N\Z),
$$
  see Fig. \ref{fig1} and \ref{fig2}. 
 \begin{figure}\lb{fig1}
\centering
\noindent
(a){
\tiny
\psfrag{g001}[l][l]{$\Gamma_{0,0,1}$}
\psfrag{g002}[l][l]{$\Gamma_{0,0,2}$}
\psfrag{g003}[l][l]{$\Gamma_{0,0,N}$}
\psfrag{g011}[c][c]{$\Gamma_{0,1,1}$}
\psfrag{g012}[c][c]{$\Gamma_{0,1,2}$}
\psfrag{g013}[c][c]{$\Gamma_{0,1,N}$}
\psfrag{g021}[c][c]{$\Gamma_{0,2,1}$}
\psfrag{g022}[c][c]{$\Gamma_{0,2,2}$}
\psfrag{g023}[c][c]{$\Gamma_{0,2,N}$}
\psfrag{g-101}[c][c]{$\Gamma_{-1,0,1}$}
\psfrag{g-102}[l][l]{$\Gamma_{-1,0,2}$}
\psfrag{g-103}[l][l]{$\Gamma_{-1,0,N}$}
\psfrag{g-111}[c][c]{$\Gamma_{-1,1,1}$}
\psfrag{g-112}[c][c]{$\Gamma_{-1,1,2}$}
\psfrag{g-113}[c][c]{$\Gamma_{-1,1,N}$}
\psfrag{g-121}[c][c]{$\Gamma_{-1,2,1}$}
\psfrag{g-122}[c][c]{$\Gamma_{-1,2,2}$}
\psfrag{g-123}[c][c]{$\Gamma_{-1,2,N}$}
\psfrag{g101}[l][l]{$\Gamma_{1,0,1}$}
\psfrag{g102}[l][l]{$\Gamma_{1,0,2}$}
\psfrag{g103}[l][l]{$\Gamma_{1,0,N}$}
\psfrag{g111}[c][c]{$\Gamma_{1,1,1}$}
\psfrag{g112}[c][c]{$\Gamma_{1,1,2}$}
\psfrag{g113}[c][c]{$\Gamma_{1,1,N}$}
\psfrag{g121}[c][c]{$\Gamma_{1,2,1}$}
\psfrag{g122}[c][c]{$\Gamma_{1,2,2}$}
\psfrag{g123}[c][c]{$\Gamma_{1,2,N}$}
\includegraphics[width=.65\textwidth,height=.5\textwidth]{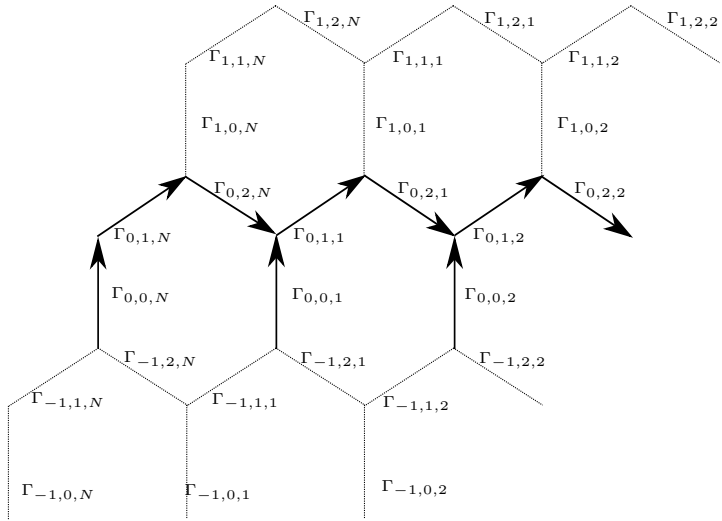}
}
(b){
\tiny
\psfrag{g-10}[l][l]{$\Gamma_{-1,0}$}
\psfrag{g-11}[l][l]{$\Gamma_{-1,1}$}
\psfrag{g-12}[l][l]{$\Gamma_{-1,2}$}
\psfrag{g00}[l][l]{$\Gamma_{0,0}$}
\psfrag{g01}[l][l]{$\Gamma_{0,1}$}
\psfrag{g02}[l][l]{$\Gamma_{0,2}$}
\psfrag{g10}[c][c]{$\Gamma_{1,0}$}
\psfrag{g11}[c][c]{$\Gamma_{1,2}$}
\psfrag{g12}[c][c]{$\Gamma_{1,3}$}
\includegraphics[height=.5\textwidth]{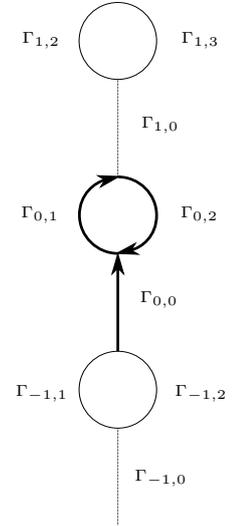}
}
\caption{(a) A piece of a nanotube $\G^N$, (b) a nanotube $\G^1$.
The fundamental domain is marked by a bold line.}
\end{figure}
\begin{figure}
\centering
\tiny
\psfrag{54}[c][c]{$-\frac{5}{4}$}
\psfrag{-1}[c][c]{$-1$}
\psfrag{1}[c][c]{$1$}
\psfrag{0}[c][c]{$0$}
\psfrag{Pi/2}[t][b]{$\frac{\pi^2}{4}$}
\psfrag{Pi}[t][b]{$\pi^2$}
\psfrag{(3*Pi)/2}[t][b]{$\frac{9\pi^2}{4}$}
\psfrag{2*Pi}[t][b]{$4\pi^2$}
\psfrag{l}[c][c]{$\l$}
\psfrag{F}[c][c]{}
\begin{tabular}{cc}
\quad\quad\includegraphics[width=.3\textwidth]{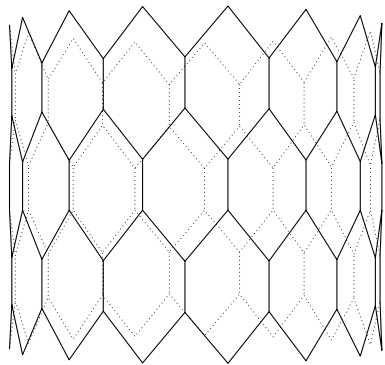}
\end{tabular}
\caption{The zigzag nanotube}
\label{fig2}
\end{figure}
Each edge $\G_\o=\{{\bf x}=
{\bf r}_\o+t{\bf e}_\o,\  t\in [0,1]\}$ is oriented by the vector ${\bf  e}_\o\in \R^3$ and has starting point $\bf r_\o\in \R^3$.
We have the coordinate ${\bf x}={\bf r}_\o+t{\bf e}_\o$ and the local coordinate $t\in [0,1]$ (length preserving).
We define ${\bf r}_{\o}, {\bf e}_{\o},\o=(n,l,j)\in \cZ$ by
${\bf e}_{n,0,j}={\bf e}_0=(0,0,1)$,
$$
{\bf e}_{n,1,j}=
{\bf \vk}_{n+2j+1}-{\bf \vk}_{n+2j}+{{\bf e}_0\/2},\qq 
{\bf e}_{n,2,j}={\bf \vk}_{n+2j+2}-{\bf \vk}_{n+2j+1}-{{\bf e}_0\/2},
\qq {\bf \vk}_j=R_N(c_{0j},s_{0j},0),
$$
$$
c_{0j}=\cos{\pi j\/N},\ s_{0j}=\sin{\pi j\/N},\ \ {\bf r}_{n,0,j}={\bf \vk}_{n+2j}+{3n\/2}{\bf e}_0,\qq {\bf r}_{n,1,j}={\bf r}_{n,0,j}+{\bf e}_0,\qq {\bf r}_{n,2,j}={\bf r}_{n+1,0,j},
$$
where $R_N={\sqrt 3\/4\sin {\pi\/2N}}$. 
The points ${\bf r}_{0,0,j}$ are vertices of the regular N-gon $\mP_0$.
The vertical edge $\G_{0,0,j}$   lies on the cylinder $\cC\ev\{{\bf x}\in \R^3:{\bf x}_1^2+{\bf x}_2^2=R_N^2\}
$. The starting points
$
{\bf r}_{1,0,j}={\bf r}_{0,2,j}=\vk_{1+2j}+{3\/2}{\bf e}_0,\ j\in \Z_N
$ 
 are the vertices of the regular N-gon $\mP_1$.
$\mP_1$ arises from $\mP_0$ by the following motion:
rotate around the axis of the cylinder $\cC$ by the angle ${\pi\/N}$
and translate by ${3\/2}{\bf e}_0$. The non-vertical vectors ${\bf e}_{0,1,j}$ and ${\bf e}_{0,2,j}$ have  positive and negative projections on the vector ${\bf e}_0$. Repeating this procedure we obtain all edges of $\G^N$. 
Note that each non-vertical edge $\G_{0,l,j}, l=1,2$ (without the endpoints) lies inside the cylinder $\cC$.
For each function $y$ on $\G^N$ we define a function $y_\o=y|_{\G_\o}, \o\in \cZ$. We identify each function $y_\o$ on $\G_\o$ with a function on $[0,1]$ by using the local coordinate $t\in [0,1]$.

Our operator $\mH_B$ on $\G^N$ acts in the Hilbert space $L^2(\G^N)=\os_\o  L^2(\G_\o)$ and is given by
\[
(\mH_B f)_\o=-\pa_{\o}^2f_\o(t)+q(t)f_\o(t),\qq
\pa_{\o}={d\/dt}-ia_\o,
\qq a_\o(t)=(\mA({\bf r}_\o+t{\bf e}_\o),{\bf e}_\o),
\]
see \cite{Ha}, \cite{SDD},
where  $(V_q f)_\o=qf_\o, q\in L^2(0,1)$  and 
$\os_\o f_\o, \os_\o f_\o'' \in L^2(\G^N)$ satisfies 

 
\no {\bf The  Kirchhoff  Magnetic  Boundary Conditions:} {\it $f$ is continuous on $\G^N$ and satisfies}
\[
\lb{KirC}
-\pa_{\o_3}f_{\o_3}(1)+\pa_{\o}f_{\o}(0)-\pa_{\o_4}f_{\o_4}(1)=0,\qqq
\pa_{\o_1}f_{\o_1}(0)-\pa_{\o}f_{\o}(1)+\pa_{\o_2}f_{\o_2}(0)=0,
\]
$$
for \ all \ 
\o_1=(n+1,0,j),\ \  \o=(n,1,j),\ \o_2=(n,2,j),\
\o_3=(n,0,j),\ \ \o_4=(n,2,j-1)\in \cZ.
$$
Condition \er{KirC} means  that  the sum of derivatives of $f$ at each vertex  of $\G^N$ equals 0 and the orientation of edges gives
the sign $\pm$.

Such models were introduced by Pauling \cite{Pa}
in 1936 to  simulate aromatic  molecules. They were described 
in more detail by  ~Ruedenberg and Scherr \cite{RS} in 1953.
For  physical models see [Ha], [SDD].

For simplicity we will denote $\G_{\o,1}\ss \G^1$ by  $\G_{\o}$, for $\o=(n,j)\in \cZ_1=\Z\ts \J$. Thus $\G^1=\cup_{\o\in \cZ_1} \G_\o$,
see Fig \ref{fig1}. The operator $\mH_B$ is unitarily equivalent to 
$H(a)=\os_1^N H_j(a),  a={3B\/16}\cot {\pi\/2N}$ (see \cite{KL1}), where the self-adjoint operator $H_j(a)$ acts in the Hilbert space $L^2(\G^1)$ and is given by  $(H_j(a) f)_\o=-f_\o''+q f_\o, f=(f_\o)_{\o\in\cZ_1}\in \gD(H_j(a))$, where $\gD(H_j(a))$ consists of all functions $f=(f_\o)_{\o\in\cZ_1}, (f_\o'')_{\o\in\cZ_1}\in L^2(\G^1)$, that satisfy  the Kirchhoff conditions
\[
\lb{1K0}
f_{n,0}(1)=f_{n,1}(0)=e^{i a}s^j f_{n,2}(1),\qq
f_{n+1,0}(0)=e^{i a}f_{n,1}(1)=f_{n,2}(0),\qq s=e^{i{2\pi \/N}},
\]
\[
\lb{1K1}
-f'_{n,0}(1)+f'_{n,1}(0)-e^{i a}s^j f'_{n,2}(1)=0,\quad
f'_{n+1,0}(0)-e^{ia}f'_{n,1}(1)+f'_{n,2}(0)=0.
\]
Define the space
$
\ell^p=\rt\{h=\{h_n\}_{1}^\iy: h_n\in\! \C\,,
\sum_{n\ge 1} |h_n|^p\!<\!\iy\rt\},\
p\ge 1$.
Let $\gS_p,p\ge 1$ be the class of conformal mappings 
$k:\C_+\to \K(h)=\{\l\in \C_+,\Re \l>0\}\sm \cup_{n\ge 1}[\pi n, \pi n+ih_n]$, where $h=(h_n)_1^\iy\in \ell^p, h_n\ge 0$
and $k(\l)=i|\l|^{1\/2}(2+O(1))$ as $\l\to -\iy$. In this case we
 introduce the sets: spectral bands $\s_n=[\l_{n-1}^+,\l_n^-]=
k^{-1}([\pi(n-1),\pi n])$ and gaps $\g_n=(\l_n^-,\l_n^+), n\ge 1$.

Recall the needed properties of the Hill operator $\wt H y=-y''+q(x)y$ on 
the real line with a periodic potential $q(x+1)=q(x),x\in \R$, see, 
e.g., \cite{MO}. We introduce the fundamental solutions $\vt(x,\l)$ 
and $\vp(x,\l),(x,\l)\in \R\ts\C$ of the equation $-y''+q(x)y=\l y$ satisfying $\vt(0,\l)=\vp'(0,\l)=1, \vt'(0,\l)=\vp(0,\l)=0$. The corresponding Lyapunov function $\D$  are given by
$\D(\l)={\vp'(1,\l)+\vt(1,\l)\/2},\ \l\in\C$.
The spectrum of $\wt H$ is purely absolutely continuous and
consists of intervals $\wt\s_n=[\wt \l_{n-1}^+,\wt \l_n^-], n\ge 1$. These intervals are separated by the gaps $\wt \g_n=(\wt \l_n^-,\wt \l_n^+)$ of length $|\wt \g_n|\ge 0$. If a gap $\wt \g_n$ is degenerate, i.e. $|\wt \g_n|=0$, then the corresponding segments $\wt\s_n,\wt\s_{n+1}$ merge. The sequence $\wt \l_0^+<\wt \l_1^-\le \wt \l_1^+\ <...$ is the spectrum of the equation $-y''+qy=\l y$ with 2-periodic boundary conditions, that is  $y(x+2)=y(x), x\in \R$.
 Here equality $\wt \l_n^-= \wt \l_n^+$ means that $\wt \l_n^\pm$ is an eigenvalue of multiplicity 2. Note that $\D(\wt \l_{n}^{\pm})=(-1)^n, \  n\ge 1$. The lowest  eigenvalue $\wt \l_0^+$ is simple, $\D(\wt \l_0^+)=1$, and the corresponding eigenfunction has period 1. The eigenfunctions corresponding to $\wt \l_n^{\pm}$ have period 1 if $n$ is even, and they are anti-periodic, that is $y(x+1)=-y(x),\ x\in \R$, if
$n$ is odd. The derivative of the Lyapunov function has a zero $\wt \l_n$ in each ''closed gap'' $[\wt \l^-_n,\wt \l^+_n]$, that is $ \D'(\wt \l_n)=0$.  Let $\m_n, n\ge 1,$ be the spectrum of the problem $-y''+qy=\l y, y(0)=y(1)=0$ (the Dirichlet spectrum). It is well-known that $\m_n \in [\wt \l^-_n,\wt \l^+_n ]$.
 Define the set $\s_D=\{\m_n, n\ge 1\}$ and note that $\s_D=\{\l\in\C: \vp(1,\l)=0\}$. Define the quasimomentum $\wt k(\l)=\arccos \D(\l)$, $\l\in \C_+$. The function
$\wt k(\cdot)\in \gS_2$, where the corresponding vector $(n^2\wt h_n)_1^\iy\in \ell^2$ is defined by the equation $\D(\wt\l_n)=(-1)^n\cosh \wt h_n$. If $\wt\l_0^+=0$, then $\wt k$ satisfies (here and below $\sqrt {-1}=i$)
\[
\lb{1}
\wt k(0)=0,\qq \wt k(\l)=z-{q_0+O(1/z)\/z},\qq \ z=\sqrt \l\in \C_+, \qq |z|\to \iy,
\]
$$
\wt k(\R_-)=i\R_+,\ 
\wt k(\wt \s_n)=[\pi(n-1),\pi n], \ \ \wt k(\wt \g_n)=[\pi n, \pi n+i\wt h_n], \ \
\wt k(\l_n)=\pi n+i\wt h_{n},n\ge 1.
$$

For a self-adjoint operator $H$ we define the set
$\s_{\iy}(H)=\{\l: \l\in\s_{pp}(H)$ is of 
infinite multiplicity$\}$. Recall needed results from \cite{KL1}.
Let $c_j=\cos a_j,\ s_j=\sin a_j, a_j=a+{\pi j\/N}$.
If $c_j\ne 0$, then the spectrum $\s(H_j(a))=\s_{\iy}(H_j(a))\cup\s_{ac}(H_j(a))$, where
$\s_{\iy}(H_j(a))=\s_D$ and 
$$
 \s_{ac}(H_j(a))=\{\l\in\R: \x_j(\l,a)\in [-1,1]\}
 =\cup_{n\ge 1}\s_{j,n}(a),
\ \ \s_{j,n}(a)=[\l_{j,n-1}^+(a),\l_{j,n}^-(a)],
$$
where $\l_{j,0}^+<\l_{j,1}^-\le \l_{j,1}^+<\l_{j,2}^-...$ are zeros of the function $\x_j^2-1$, and 
 $\x_j$ is  the {\bf modified Lyapunov functions} given by
\[
\lb{eilf}
\x_j={F+s_j^2\/c_j}, \qqq    F={9\D^2-\D_-^2-5\/4},\ \qq {\rm where} \qq \D_-={\vp'(1,\cdot)-\vt(1,\cdot)\/2},
\qq   j\in \Z_N.
\]
If $c_j=0$, then the spectrum $\s(H_j(a))=\s_{\iy}(H_j(a))=\s_D \cup
\{\l\in \R: F(\l)=-1\}$.

If $\l\in \s_{ac}(H_j(a))$, then the equation $-y''+qy=\l y$
on $\G^1$ with conditions \er{1K0},\er{1K1} has a solution
$\p$ such that $\p_{n+1,0}(0)=e^{ip_j(\l)}\p_{n,0}(0), \p_{n+1,0}'(0)=e^{ip_j(\l)}\p_{n,0}'(0), n\in \Z$, where $p_j(\l)$ is a quasimonentum. The function $\cos p_j(\l),j\ne 0$ is not entire, and it is define on some Riemann surface. If we take
$k_j=p_j+{\pi j\/N}$, then $\x_j=\cos k_j(\l)$
is the entire function.

Below we consider only the operator $H_0(a), c=\cos a>0$.
Due to \er{eilf}, the results for the operator $H_0(a+{\pi j\/N})$  give the results for the operator $H_j(a)$.
 
For simplicity we will write $\x=\x_0$ and $\l_{n}^{\pm}=\l_{n}^{\pm}(a)=\l_{0,n}^{\pm}(a),n\ge 0$.
The function  $\x^2-1$ has only real zeros
$\l_{n}^{\pm}$, their labeling is given by
$\l_{0}^{+}<\l_{1}^{-}\le \l_{1}^{+}<\l_{2}^{-}\le \l_{2}^{+}<..$ and if $|c|\notin \{{1\/2},1\}$, then $\l_{n}^-<\l_{n}^+$. Here
$\s_n=[\l_{n-1}^+,\l_{n}^-]$ are the spectral bands and $\g_n=(\l_{n}^-,\l_{n}^+)$ are the gaps. Moreover,
 $\l_{n}^{\pm}$ satisfy $\x(\l_{n}^{\pm})=(-1)^n$  and  
\begin{multline}
\lb{as-pe}
\l_{n}^{\pm}=\l_{n}^{0,\pm}+q_0+\ve_n^\pm\qq   
\ \qq as \ \ n\to \iy,\qq q_0=\int_0^1q(t)dt,
\\
\ve_{n}^\pm=\pm |\wt q_{cn}|+o(n^{-1}),
\qq \wt q_{cn}=\int_0^1q(t)\cos \pi nt dt,
\qqq  n\ is \ odd, \qq c={1\/2}
\\
\ve_{n}^{\pm}=\pm\rt|{|\hat q_n|^2}-{\hat q_{sn}^2\/9}\rt|^{1\/2}+{O(1)\/n},\qq \hat q_n=\int_0^1q(t)e^{i2\pi nt}dt,\qq \hat q_{sn}=\Im \hat q_{n},\
n\ is \ even,\qq c=1,
\\
\ve_n^\pm=o(1/n) \qq in\ other \ cases,
\end{multline}
see \cite{KL}, \cite{KL1}.
Here $\l_{n}^{0,\pm}$ are 2-periodic eigenvalues
for the case $q=0$ given by
\begin{multline}
\lb{as-pe0}
\sqrt{\l_0^{0,+}}=\f_0\in [0,\pi/2],\qq \sqrt{\l_{2n}^{0,\pm}}=\pi n\pm \f_0,\qq 
\cos 2\f_0={8\/9}\rt(c^2+c-{7\/8}\rt)\in [-1,1],\\
\sqrt{\l_{2n+1}^{0,\pm}}=\pi (n+{1\/2})\pm \f_1,\qq \f_1\in [0,\pi/2], \qq 
-\cos 2\f_1={8\/9}\rt(c^2-c-{7\/8}\rt)\in [-1,1].
\end{multline}

The identity $\x(\l)=\cos k(\l), \l\in \C_+$ defines an analytic 
function (the quasimomentum) $k(\l),\l\in \C_+$, see Theorem \ref{T2}.
With  each edge  of  the gap $\g_n\ne \es$, we associate the effective mass $\m_0^+, \m_n^{\pm}$ by (we take some branches $k$ such that $k(\l)\to \pi n$ as $\l\to \l_n^{\pm}$) 
\[
\l=\l_n^{\pm}+{(k(\l)-\pi n)^2\/2\m_n^{\pm}}(1 +o(1))
\ \ \ \ \  {\rm as} \ \ \ \             \l\to \l_n^{\pm},
\]
and let $\m_n^{\pm}=0$ if $|\g_n|=0$.
If $q=0$, then the effective masses $\m_{n}^{0,\pm}$ are given by (see Sect. 2)
\[
\lb{em0}
\m_0^{0,+}={9\/8c}{\sin 2\f_0\/\f_0},\qq
\m_{2n}^{0,\pm}=\pm {9\/8c}{\sin 2\f_0\/\sqrt{\l_{2n}^{0,\pm}}},\qq
\m_{2n-1}^{0,\pm}=\pm {9\/8c}{\sin 2\f_1\/\sqrt{\l_{2n-1}^{0,\pm}}},\qq
n\ge 1,
\]
and $\m_n^{0,+}+\m_n^{0,-}=O(1/n^2)$ as $n\to\iy$.  Let $F_0={9\cos 2\sqrt{\l}-1\/8}$.
We formulate our first result.

\begin{theorem} \label{T1} 
Let $q\in L^2(0,1)$. Then the following identities and asymptotics hold true
\[
\lb{T1-1}
\m_n^\pm=\m_n^{0,\pm}+{(-1)^{n+1}F_0''(\l_n^{0,\pm})\ve_n^\pm\/c}+{O(1)\/n^3}\qqq as \qqq n\to \iy,
\]
\[
\lb{T1-2}
k'(\l)^2={1\/2}\sum _{n\ge 0, \n=\pm}{ \mu_n^\n \/\l-\l_n^\n},\qq
\l\ne \l_n^\pm,\ n\ge 0,
\]
\[
\lb{T1-3}
\m_0^++\sum_{n\ge 1}(\m_n^++\m_n^-)=2,\
\]
\[
\lb{T1-4}
\m_{2n}^{\pm}=
2\sum_{m\ge 1, s=\pm}(\l_{2m-1}^s-\l_{2n}^{\pm})^{-1},\ \ \ \
\m_{2n+1}^{\pm}=2\sum_{m\ge 0, s=\pm}(\l_{2m}^s-\l_{2n+1}^{\pm})^{-1}, \
n\ge 0,
\]
where the series converges absolutely and uniformly on compact sets in $\C\sm\{\l_n^\pm,\ n\ge 0\}$.
\end{theorem} 

{\bf Remark.} Note that $F_0''(\l_n^{0,\pm})\ve_n^\pm=O(\ve_n^\pm/n^2)$
as $n\to \iy$.

Let $\l_n, n\ge 1$ be the zeros of $F'$. We have $\x'(\l_n)=0$.
Recall that $q_0=\int_0^1q(x)dx$.

\begin{theorem} \label{T2} 
i) Let $q\in L^2(0,1)$ and $\l_0^+=0$. Then
a quasimomentum $k(\l)=\arccos \x(\l)$, $\l\in \C_+$ belongs to
$\gS_\iy$, $(h_n)_1^\iy$ is defined by the equation $\x(\l_n)=(-1)^n\cosh h_n$,  and satisfies
\[
\lb{T2-1}
k(0)=0,\qq k(\l)=2z+\log \rt({9\/8c}\rt)^2-{q_0+O(1/z)\/z},\qq z=i|\l|^{1\/2},\ \l=z^2\to -\iy,
\]
\[
\lb{T2-2}
k(\R_-)=i\R_+,\ 
k(\s_n)=\![\pi(n-1),\pi n], \  k(\g_n)=\![\pi n,\pi n+ih_n], \ 
k(\l_n)\!=\!\pi n+ih_{n},n\ge1.
\]
ii) Moreover, for each $n\ge 1$ the following estimates hold true
\[
\lb{T2-3}
h_n \le 3\pi \sqrt{|\g_n||\m_n^{\pm }|\/2}\le 6\pi^2n(\m_n^+-\m_n^-),
\]
\[
\lb{T2-4}
|\g_n| \le  (4\pi n)^2(\m_n^+-\m_n^-),
\]
\[
\lb{T2-5}
|\g_n|\le 8\l_n^+\m_n^+,\qq |\g_n|\le 8\l_n^-|\m_n^-|+16\l_n^-(\m_n^-)^2,
\]
\[
\lb{T2-6}
h_n\le  4\pi \sqrt{\l_n^\pm}|\m_n^\pm|,\qq
h_n\le  \pi\sqrt{2|\g_n||\m_n^-|},\qqq  
h_n\le 2\pi\sqrt{|\g_n|\m_n^+},
\]
\[
\lb{T2-7}
h_n^2\le 2|\g_n| \sqrt {\m_n^+|\m_n^-|}.
\]
If $\g_n$ is the first non degenerate gap for some $n\ge 1 $, then $\m_0^+\ge -\m_n^-$. 

\no iii) Moreover, let a spectral interval  $\s(n,n_1)=[\l_{n}^+,\l_{n_1}^-]=\cup_{n+1}^{n_1}\s_j$, where
$n_1-n$ is a number of the merged components $\s_j$ which are composed this interval $\s(n,n_1)$. Then
\[
\lb{T3-3}
\m_{n}^+(\l_{n_1}^--\l_{n}^+)\le 16(n_1-n)^2(\l_{n}^++\l_{n_1}^-),\qqq  |\m_{n_1}^-|(\l_{n_1}^--\l_{n}^+)\le 32(n_1-n)^2\l_{n_1}^-.
\]

\end{theorem}

{\bf Remark.} 1) There is a big difference beween
the quasimomentum for the operator $H_0$ and the Hill operator
$\wt H$: $h\in \ell^\iy$ and $(n^2\wt h_n)_1^\iy\in \ell^2$,
and, in particular, the  unperturbed vector $h^0\in \ell^\iy$.
2) Note that $n_1-n\le 2$ in \er{T3-3}.

Below we  will sometimes write $\l_n^\pm(a), \m_n^\pm(a), \x(\l,a),..$, instead of $\l_n^\pm, \m_n^\pm, \x(\l),..$, when several magnetic fields are being dealt with.

\begin{theorem}\label{T3} 
Let $q\in L^2(0,2)$ and $a\in [0,{\pi\/2}), n\ge 0$. Then
\[
\lb{T3-1}
h_n^{0}(a)\le h_n(a),\qq
|\m_n^{0,\pm}(a)|\le|\m_n^{\pm}(a)|,\qqq
|\s_n^0(a)|\ge |\s_n(a)|,
\]
\[
\lb{T3-2}
h_n(a)\le h_n(a_1),\qq
|\m_n^{\pm}(a)|\le|\m_n^{\pm}(a_1)|,\qq
|\s_n^0(a)|\ge |\s_n(a_1)|,
\qq all \qq {\pi\/3}\le a\le a_1\le {\pi\/2}.
\]
\end{theorem}

In the present paper we obtain only local estimates, i.e., 
estimates for fix $n\ge 0$. For the Hill operator there exist
a priori two sided estimates \cite{K1}, \cite{K2}. In order
to obtain similar results for the zigzag nanotubes, where $h\in \ell^\iy$,  we have to study carefully the
quasimomentum $k(\cdot)$ as  conformal mapping. In our paper
we only touch this problem.

\section {Identities and asymptotics}
\setcounter{equation}{0}

Recall that $F={9\D^2-\D_-^2-5\/4}$, where $\D_-={\vp'(1,\cdot)-\vt(1,\cdot)\/2}$
 and $\D, F$ satisfy $q_0=\int_0^1q(t)dt,$
\[
\label{Das} \D(\l)= \cos\sqrt{\l}+
{q_0\sin \sqrt{\l}\/2\sqrt{\l}}+{O(e^{|\Im\sqrt{\l}|})\/|\l|},\qq \qqq
\D_-(\l)={o(e^{|\Im\sqrt{\l}|})\/|\l|^{1\/2}},
\]
\[ 
\lb{asD0}
 F(\l)=F_0(\l)+{9q_0\/8}{\sin 2\sqrt{\l}\/\sqrt{\l}}+O\lt({e^{2|\Im\sqrt{\l}|}\/|\l|}\rt),\ \ 
F_0(\l)={9\cos 2\sqrt{\l}-1\/8},
\]
\[ 
\lb{asD1}
 F'(\l)=F_0'(\l)+{9q_0\/8}{\cos 2\sqrt{\l}\/\l}+O\lt({e^{2|\Im\sqrt{\l}|}\/|\l|^{3\/2}}\rt)
\]
as $|\l |\to\iy$, uniformly on bounded sets of $q\in L^2(0,1)$ 
 (see [KL]).

The identity $\m_n^\pm=-(-1)^n\x'(\l_n^\pm),n\ge 0$
from \cite{KK1} together with $\x={F+s^2\/c}$ yields
\[
\lb{demkk}
\m_n^\pm=-(-1)^n\x'(\l_n^\pm)=-(-1)^n{F'(\l_n^\pm)\/c},\qq n\ge 0.
\]

{\bf In the unperturbed case } $q=0$  {\bf the modified Lyapunov function} is given by
\[
 \x^0={F_0+s^2\/c}, 
\qq F_0={9\cos 2z-1\/8}\qqq  c=\cos a>0,\ \  s=\sin a,\qq z=\sqrt \l.
\]
The function $\x^0(z^2)$ is $\pi$-periodic and on the period $[0,\pi]$
has a maximum at $z\in {0,\pi}$ and a minimun at $z={\pi\/2}$:
\[
\max_{x\in \R} \x^0(x^2)=\x(0)={1+s^2\/c}={2-c^2\/c}>1\ \ if \ \ c\ne 1
\qq and \qq \max_{x\in \R} \x^0(x^2)=1 \ if \ \ c=1,
\]
\[
\min_{x\in \R} \x^0(x^2)=\x(\pi^2/4)=-c-{1\/4c}<-1\ \ if \ \ 
c\ne {1\/2}
\qq and \qq \x^0(\pi/2)=-1 \qq if \ c={1\/2}.
\]
We have gaps $\g_n^0=(\l_n^{0,-},\l_n^{0,+})$, where $\x^0(\l_n^{0,\pm})=(-1)^n$.  Using $\x^0(\l_n^{0,\pm})=(-1)^n$ we have 
the following equetion for $z_n^{0,\pm}=\sqrt{\l_n^{0,\pm}}>0$:
$$
9\cos 2z-1+8s^2=8c(-1)^n,\qq \cos 2z 
={8\/9}\rt(c^2+c(-1)^n-{7\/8}\rt)\in [-1,1].
$$
Then 2-periodic eigenvalues $\l_n^{0,\pm}=(z_n^{0,\pm})^2$ have the form
\er{as-pe0}, i.e., 
$$
z_0^{0,+}=\f_0\in [0,{\pi\/2}],\qq z_{2n}^{0,\pm}=\pi n\pm \f_0,\qq 
\cos 2\f_0={8\/9}\rt(c^2+c-{7\/8}\rt)\in [-1,1],
$$
$$
z_{2n-1}^{0,\pm}=\pi (n-{1\/2})\pm \f_1,\qq \f_1\in [0,\pi/2], \qq 
\cos 2({\pi\/2}-\f_1)=-\cos 2\f_1={8\/9}\rt(c^2-c-{7\/8}\rt)\in [-1,1]
$$
for $ n\ge 1$. The function $k^{0}=\arccos \x^0(\l)$ is a conformal mapping from the upper half-plane $\C_+$ onto a quasimomentum domain 
$\K(h^0)=\C\sm \cup_{n\ge 1} [\pi n, \pi n+ih_n^0]$, where $h=(h_n^0)_1^\iy, h_n\ge 0$
is defined by the equation $\cosh h_n^0=(-1)^n\x(\l_n)\ge 1$ and 
satisfies
\[
\cosh h_0^0={1+s^2\/c}\ge 1,\qq h_{2n}^0=h_{0}^0,\qq and \qq
\cosh h_1^0={1+4c^2\/4c}\ge 1, \qq h_{2n+1}^0=h_{1}^0.
\]
Using \er{demkk}, we deduce that  {\bf effective masses} for $q=0$ are given by
\[
\m_n^{0,\pm}=(-1)^{n+1}{F'(\l_n^{0,\pm})\/c}={9(-1)^n\/8c}{\sin 2z_n^{0,\pm}\/z_n^{0,\pm}},\qq \l_n^{0,\pm}=(z_n^{0,\pm})^2,\qq
n\ge 0.
\]
Thus
$$
\m_0^{0,+}={9\/8c}{\sin 2\f_0\/\f_0},\qq
\m_{n}^{0,\pm}=\pm {9\/8c}{\sin 2\f_0\/({\pi n\/2}\pm \f_0)},\qq
\m_{n}^{0,+}+\m_{n}^{0,-}=-{9\f_0\sin 2\f_0\/c4[{(\pi n)^2\/4}-\f_0^2]}<0,\qq
n \ \ is \ even,
$$
and 
$$
\m_n^{0,\pm}=\pm {9\/8c}{\sin 2\f_1\/({\pi n\/2}\pm \f_1)},\qq
\m_n^{0,+}+\m_n^{0,-}=-{9\f_1\sin 2\f_1\/c4({(\pi n)^2\/4} -\f_1^2)}<0,\qq 
n \ \ is \ odd.
$$

\no{\bf Proof of Theorem \ref{T1}.} 
Using asymptotics \er{as-pe} we obtain
$$
F'(\l_n^\pm)=F_0'(\l_n^\pm)+O(1/n^3),\ \ 
F_0'(\l_n^\pm)=F_0'(\l_n^{0,\pm})+F_0''(\l_n^{0,\pm})\ve_n^\pm+O(1/n^3)
$$
as $n\to\iy$. Combine  last asymptotics with  $\m_n^\pm=-(-1)^n{F'(\l_n^\pm)\/c}$ (see \er{demkk}) we get \er{T1-1}.

 Using the Cauchy theorem about residues for the function
 ${k'}^2={{\x'}^2\/\x^2-1}$, we deduce that
$$
{1\/2\pi i}\int_{|\r|=t}{k'(\r)^2\/\r-\l}d\r=k'(\l)^2-{1\/2}\sum _{|\l_n^\n|<t, \n=\pm}{ \mu_n^\n \/\l-\l_n^\n},\qqq all \ \ t\ne \l_n^\pm, n\ge 0.
$$
The identity $\x={F+s^2\/c}$ and $b_\pm=\pm c-s^2$  gives
\[
\lb{eik12}
{k'}^2={{\x'}^2\/1-\x^2}={{F'}^2\/c^2(c^2-(F+s^2)^2}
={-{F'}^2\/c^2(F-b_-)(F-b_+)}.
\]
Using $\cos 2\f_\pm={1+8b_\pm\/9}\in [-1,1],\qq \f_\pm\in [0,{\pi\/2}]$, we obtain  
\[
\lb{31}
F_0(\r)-b_\pm={9\/8}(\cos 2z-\cos 2\f_\pm)=-{9\/4}\sin (z-\f_\pm)
\sin (z+\f_\pm),\qq z=\sqrt \r.
\]
We need the simple estimates (for each $r\in (0,{\pi\/2}]$)
\[
\lb{ses}
2|\sin z|\ge e^{|\Im z|}(1-e^{-2r}), \qq any \qq z\in \cC_r=\{z\in \C: |z-\pi n|\ge r, n\in \Z\}.
\]
 Using \er{31},\er{ses} we obtain for $\z_1=z-\f_\pm,\ \ \z_2=z+\f_\pm\in \cC_r$ (we take small $r<<1$)
\[
\lb{ses1}
|F_0(\r)-b_\pm|={9\/4}|\sin \z_1
\sin \z_2|\ge {9\/4}e^{2|\Im z|}(1-e^{-2r})^2\qq 
as \qq |\r|\to \iy.
\]
Substituting \er{asD0},\er{asD1} and \er{ses1}   into \er{eik12} we deduce that
$$
k'(\r)^2={-{F'(\r)}^2\/c^2(F(\r)-b_-)(F(\r)-b_+)}={O(\r^{-1})e^{4|\Im \r|}\/e^{4|\Im \r|}(1+o(1))}=O(\r^{-1})\qq as \qq t\to \iy,
$$
for all $\r\in \{\m\in \C: |\m|=t\}\ss\cC_r$, where $\sqrt t-\f_\pm,
 \sqrt t-\f_\pm\in \cC_r$. Thus we obtain
 $\int_{|\r|=t}{k'(\r)^2\/\r-\l}d\r=O(1/n)$ as $t\to \iy$,
 which yields \er{T1-2}.

In order to show \er{T1-3} we need the following identities
$$
{ \m_n^+ \/\l-\l_n^+}+{\m_n^- \/\l-\l_n^-}=A_n+B_n,\qqq
A_n={(\m_n^++\m_n^-)\/2}\rt({1\/\l-\l_n^+}+{1\/\l-\l_n^-}\rt),
$$$$
B_n={(\m_n^+-\m_n^-)\/2}\rt({1\/\l-\l_n^+}-{1\/\l-\l_n^-}\rt)=
{(\m_n^+-\m_n^-)(\ve_n^+-\ve_n^-)\/2(\l-\l_n^+)(\l-\l_n^-)}. 
$$
Asymptotics \er{T1-1} give $\sum |\m_n^++\m_n^-|<\iy$ and
$\sum |(\m_n^+-\m_n^-)(\ve_n^+-\ve_n^-)|<\iy$, which implies
$$
\l\rt({ \m_0^+ \/\l-\l_0^+}+\sum_{n\ge 1}A_n\rt)\to \m_0^++
\sum_{n\ge 1}(\m_n^++\m_n^-),\qq and \qq \l\sum_{n\ge 1}B_n\to 0
$$
as $\l\to -\iy$. 
These asymptotics together with \er{T1-2}, \er{ak'2-1} yield \er{T1-3}.

The Hadamard factorization
$\x(\l)+1\!\!=\!\!\!\!\prod\limits_{n\ge 0,s=\pm}\lt(1-{\l\/ \l_{2n+1}^s}\rt)$
gives ${\x'(\l)\/\x(\l)+1}=\!\!\!\!\sum\limits_{n\ge 0,s=\pm}{1\/\l-\l_{2n+1}^s}$. Then the identity \er{demkk}  implies
$$
\m_{2n}^\pm=-\x'(\l_{2n}^\pm)=-{2\x'(\l_{2n}^\pm)\/\x(\l_{2n}^\pm)+1}=2\sum_{n\ge 0,\n=\pm}{1\/\l_{2n}^\pm-\l_{2n+1}^\n},
$$
which yields \er{T1-4}. The proof for $\m_{2n+1}^\pm$ is similar.
\BBox

\section {Conformal mappings and estimates}
\setcounter{equation}{0}

{\bf Proof of Theorem \ref{T2}. } i) We need some results from \cite{MO}.
Let a function $f$ be entire and $f(z^2), z\in \C$ of exponential type 2 and $f(\l)$ be real on the real line
and $f(\l)=O(1)$  and $f(-\l)=C_fe^{2|\l|^{1\/2}+o(1)}$ as $\l\to+\iy$ for some constant $C_f>0$. Assume that all zeros of the function $f^2-1$ are real and their labeling is given by
$\z_{0}^{+}<\z_{1}^{-}\le \z_{1}^{+}<\z_{2}^{-}\le \z_{2}^{+}<..$..
Then there exists a conformal mapping $k:\C_+\to \K(h)$
for some sequence $h=(h_n)_1^\iy\in \ell^\iy$ such that
$f(\l)=\cos k(\l)$ and $k(\l)=\sqrt{\l}(2+O(1))$ as $\l\to -\iy$
and if $\z_{0}^{+}=0$, then  $k$ satisfies
 $$
k(\R_-)=i\R_+,\ 
k([\z_{n-1}^+,\z_{n}^-])=[\pi(n-1),\pi n], \ 
k([\z_{n}^-,\z_{n}^+])=[\pi n, \pi n+ih_n],\ 
k(\z_n)=\pi n+ih_{n}, 
$$
$n\ge 1$, where $\z_1<\z_2<\z_3<..$ are zeros of $f$ and $\z_n\in [\z_{n}^-,\z_{n}^+]$ for all $n\ge 1$.

The function $\x$ satisfies these conditions, then 
 the statement i) have been proved and we need only to show 
\er{T2-1}. Identities $\D(\l)=\cos \wt k(\l)$ and asymptotics
$\wt k(\l)=z-{q_0+o(1)\/2z}, z=iy=\sqrt\l, y\to\iy$, see \cite{MO}, and  \er{Das}  give
$$
F(\l)={9\/8}(\cos 2\wt k(\l)+o(\l^{-1}e^{2y})),\qq
\x(\l)={9\/8c}\cos 2\wt k(\l)+o(z^{-1}e^{2y})={9\/16c}e^{-2i\wt k(\l)
+o(z^{-1})}.
$$
Then $\cos k(\l)={e^{-ik(\l)}\/2}(1+O(e^{-4y}))$
 yields \er{T2-1}.

ii) We have proved the existence of the conformal mapping
$k:\C_+\to \K(h)$ for some $h\in \ell^\iy$. For such conformal mapping the estimates \er{T2-3}, \er{T2-4} were proved in \cite{K1}.
Moreover, if $\g_n$ is the first
non degenerate gap for some $n\ge 1 $, then $\m_0^+ \ge -\m_n^-$
\cite{K1}.

Let $n\ge 1$. We need the following estimates from \cite{KK} 
\[
\lb{T2-33}
{|g_n|\/2}\le h_n\le \pi\sqrt{2|g_n||m_n^{\pm}|} 
\le 2\pi |m_n^{\pm}|,\qqq where \qq 2z_n^{\pm}m_n^{\pm}=\m_n^{\pm},\qq z_n^{\pm}=\sqrt{\l_n^{\pm}}>0
\]
\[
\lb{T2-44}
h_n^2\le 2|g_n| \sqrt {m_n^+|m_n^-|},\qqq
|g_n|\le 2|m_n^{\pm}|,\qqq \qqq where \qq (z_n^++z_n^-)|g_n|=|\g_n|.
\]
Consider $\m_n^+$. Using the estimate $|g_n|\le 2|m_n^+|$
we obtain
$$
|\g_n|\le 4(z_n^++z_n^-)z_n^+\m_n^+\le 8\l_n^+\m_n^+,
$$
which yields the first estimate in \er{T2-5}. Consider $\m_n^-$. Using
the estimate $|g_n|\le 2|m_n^-|$ and identites from \er{T2-33}, \er{T2-44} we obtain
$$
|\g_n|\le 4(z_n^++z_n^-)z_n^-\m_n^-\le 8\l_n^-\m_n^-+4|g_n|z_n^-\m_n^-,
\qq |g_n|z_n^-\le 4 \l_n^-\m_n^-,
$$
which yields \er{T2-5}. 

The estimate from \er{T2-33} givees
$h_n\le 2\pi |m_n^{\pm}|=4\pi \sqrt{\l_n^{\pm}}|\m_n^{\pm}|$,
which gives the first estimate in \er{T2-6}.
Using estimates  
$$
|g_n||m_n^-|\le |\g_n||\m_n^-|,\qqq |g_n|m_n^+\le 2|\g_n|\m_n^+
$$
and \er{T2-33}, \er{T2-44}, we deduce that  
$$
h_n\le  \pi\sqrt{2|\g_n||\m_n^-|},\qqq  
h_n\le 2\pi\sqrt{|\g_n|\m_n^+},
$$
which yields the last two estimates in \er{T2-6}.
The identities from \er{T2-33}, \er{T2-44} give
$$
|g_n|^2m_n^+|m_n^-|=4z_n^+z_n^-|g_n|^2\m_n^+|\m_n^-|\le |\g_n|^2\m_n^+|\m_n^-|
$$
since $4z_n^+z_n^-\le (z_n^++z_n^-)^2$. This implies \er{T2-7}.

iii) We need the estimate  (see Theorem 2.4 from \cite{KK})
$$
|\m_{\pm}||\s(n,n_1)|\le 16(n_1-n)^2\sqrt{\l_{\pm}}(\sqrt{\l_-}+\sqrt{\l_+}),
\qq \m_+=\m_{n}^+, \l_+=\l_{n}^+,
\m_-=\m_{n_1}^-, \l_-=\l_{n_1}^-,
$$
 where $n_1-n$ is the number of the merged components which are
composed the band $(\l_+,\l_-)$. This estimate and $\l_+<\l_-$ yields \er{T3-3}.
\BBox

{\bf Proof of Theorem \ref{T3}. }
Using the estimates $|F(\l_n)|\ge |F_0(\l_n^0)|\ge 1$ (see Lemma 3.1 from \cite{KL}) we deduce that $h_{n}^0\le h_{n}$ for all $n\ge 1$.
Applying these facts to the qusimomentums $k,k^0$ and using \er{esq} we get \er{T3-1}.

We show \er{T3-2}.
If $n$ is even, then  $F(\l_n)\ge 1$ and $f_n(c)=\cosh h_n={F(\l_n)+s^2\/c}=-c+{F(\l_n)+1\/c}$ and
$f_n'(c)=-1-{F(\l_n)+1\/c^2}<-1$.

If $n$ is odd, then $F(\l_n)\le -{5\/4}$ (see Lemma 3.1 from \cite{KL}) and  $f_n(c)=\cosh h_n=-{F(\l_n)+s^2\/c}=c-{F(\l_n)+1\/c}$ and
$f_n'(c)=1+{F(\l_n)+1\/c^2}$. Assume that $c^2<{1\/2}$, then
$f_n'(c)\le{c^2-{1\/4}\/c^2}<0$.
Thus the function $h_n(a)$ on the interval $[{\pi\/3},{\pi\/2}]$ is increasing and $h_n(a)<h_n(a_1)$ for all ${\pi\/3}\le a<a_1\le {\pi\/2}$. Then estimate \er{esq} yields \er{T3-2}.
\BBox

\begin{lemma}
\lb{ak'2}
The  asymptotics \er{T2-1} and  following one hold true
\[
\lb{ak'2-1}
k'(\l)^2={1\/\l}+{q_0+o(1)\/\l^2}
\qqq \qqq as \qq \l\to -\iy.
\]
\end{lemma}
\no{\bf Proof.} Using \er{Das}-\er{asD1}, we obtain
$$
k'(\l)^2=-{\x'(\l)^2\/\x(\l)^2-1}=-{\x'(\l)^2\/\x(\l)^2}(1+O(e^{2y})),
\qqq 
{\x'(\l)\/\x(\l)}={F'(\l)\/F(\l)}(1+O(e^{2y})),
$$
where $y=\sqrt{-\l}>0$. Identities $\D(\l)=\cos \wt k(\l), z=iy=\sqrt\l$ and asymptotics $\wt k(\l)=z-{q_0+o(1)\/2z},\ \ \wt k'(\l)={1\/2z}(1+{q_0+o(1)\/2\l})$, (see \cite{MO})
and  \er{Das}  give
$$
F(\l)={9\/8}(\cos 2\wt k(\l)+o(\l^{-1}e^{2y})),\qq
F'(\l)={9\/8}(-\sin 2\wt k(\l)) 2 \wt k'(\l)    +o(z^{-{3}}e^{2y}),
$$
$$
{F'(\l)\/F(\l)}=-{\sin \wt k(\l)\/\cos \wt k(\l)}(2\wt k'(\l)+o(\l^{-{1}}))=
i(2\wt k'(\l)+o(z^{-3})={i\/z}(1+{q_0+o(1)\/2\l}),
$$
which yields \er{ak'2-1}.
\BBox

\begin{lemma}
\lb{esq}
Let $k_1,k_2\in \gS_\iy$ and let $h_{1,n}\le h_{2,n}$
for all $n\ge 1$. Then the corresponding spectral bands 
$\s_{1,n},\s_{2,n}$ and effective
masses $\m_{1,n}^{\pm}, \m_{2,n}^{\pm}$ satisfy
\[
\lb{esq}
|\s_{1,n}|\ge |\s_{2,n}|,   \qqq
|\m_{1,n}^{\pm}|\le |\m_{2,n}^{\pm}|,\qq all \qqq n\ge 1.
\]
\end{lemma}
\no{\bf Proof.} Let $r_j(\l)=k_j^2(\l), \l=\z+i\e\in \C_+,j=1,2$, where $k_j=u_j+iv_j$.
The function $r_j$ is a conformal mapping from $\C_+$
onto $\cR_j=\{r=k^2, k\in \K(h_j)\}$. Let $\l_j(r),r\in \cR_j$ be the inverse mapping $\l_j=r_j^{-1}$.  
The function $s_j=\Im r_j=2u_jv_j$ is harmonic, nonnegative in $\C_+$ and $s_j\in C(\ol\C_+)$ and $r_j$ satisfies
\[
\lb{repr}
r_j(\l)=\l+C_j+{1\/\pi}\int_{\l_{j,1}^-}^{\iy }\!\! s_j(t)\lt( {1\/t-\l} -
{t\/1+t^2}\rt)dt, \qqq C_j=-{1\/\pi}\int_{\l_{j,1}^-}^{\iy }{s_j(t)dt\/t(1+t^2)}, \ \l\in \C_+,
\]
where $\l_{j,1}^->0$ and $\int_{\l_{j,1}^-}^{\iy }{s_j(t)dt\/(1+t^2)}<\iy$,
see \cite{K1}. In the domain 
$D_\ve=\{\l\in \C_+: \ve\le \arg \l\le \pi-\ve\},0<\ve<{\pi\/2}$
there is an estimate $|t-\l|\ge |t|\sin \ve$ for all $t>0$.
This and \er{repr} yields $r_j(\l)=\l(+o(1))$ as $\l\in D_\ve, |\l|\to\iy$.
  But for any  $\vk$ there exists a constant $\r=\r(\vk)>0$ such that
$\{\l:|\l|>\r\}\cap D_\vk\ss r_j(D_\ve),\ \ j=1,2$ for some $\ve<\vk<{\pi\/2}$.
Then $\l_j(r)=r(1+o(1)),\ \  r=t+is\in D_\vk$ as $|r|\to\iy$, and
$$
{\l_1(r_2(i\e))\/i\e}={\l_1(r_2(i\e))\/(r_2(i\e))}{r_2(i\e))\/i\e}
\to 1 \ \ {\rm as}\ \ \e\to\iy,
$$
which yields   $f(i\e)=\e(1+o(1))$ as $\e\to\iy$. Then the Herglotz Theorem yields
\[
\lb{apes1}
f(\l)={\rm Im} \l_1(r_2(\l))\ge  {\rm Im} \l_2(r_2(\l))=\e, \qqq \l=\z +i\e\in \C_+.
\]
Then $\Im\l_1\ge\Im\l_2\ge 0$ in the domain $r_2(\C_+)$ and 
$\Im\l_1(r)=\Im\l_2(r)=0, r\in \R$ give
$$
\l_1'(r)={\pa\/\pa s} {\rm Im} \l_1(r) \ge
{\pa\/\pa s} {\rm Im} \l_2(r)=\l_2'(r), \ \ r\in\R,
r\neq t_n=(\pi n)^2,
$$
(recall that $r=t+is$) which implies
$$
|\s_{1,n}|=\int _{t_{n-1}}^{t_n}\l'_1(r)dr \ge
\int _{t_{n-1}}^{t_n}\l'_2(r)dr=|\s_{2,n}|, \ \ \ n\ge 1.
$$
Moreover, we deduce that
$$
  \l_j(r)-\l_{j,n}^+=\int_{t_n}^r\l_j'(r)dr
={(r-t_n)^2\/(2\pi n)^22\m_{j,n}^+}(1+o(1))\qq as\ \
 r\to t_n+0,
$$
which yelds $\m_{1,n}^+\le \m_{2,n}^+$. The proof for
$\m_{1,n}^-$ is similar.
  \BBox

\end{document}